# Statistical effects of dose deposition in track-structure modelling of radiobiology efficiency

M. Beuve[1], A. Colliaux[1], D. Dabli[2], D. Dauvergne[1], B. Gervais[3], G. Montarou[2], E.Testa[1]

[1]*Université de Lyon, F-69622, Lyon, France; Université Lyon 1, Villeurbanne; CNRS/IN2P3, UMR5822, Institut de Physique Nucléaire de Lyon ou IPNL;*

[2]*Laboratoire de Physique Corpusculaire de Clermont-Ferrand, IN2P3, Université Blaise Pascal, Clermont-Ferrand, France*

[3]*CIMAP, CEA, CNRS, ENSICAEN, UCBN, Caen France*

**Abstract: Ion-induced cell killing has been reported to depend on the irradiation dose but also on the projectile parameters. In this paper we focus on two approaches developed and extensively used to predict cell survival in response to ion irradiation: the Local Effect Model and the Katz Model. These models are based on a track-structure description summarized in the concept of radial dose. This latter is sensitive to ion characteristics**

**parameters and gives to both models the ability to predict some important radiobiological features for ion irradiations. Radial dose is however an average quantity, which does not include stochastic effects. These radiation-intrinsic effects are investigated by means of a Monte-Carlo simulation of dose deposition. We show that both models are not fully consistent with the nanometric and microscopic dose deposition statistics.**

**A INTRODUCTION**

Cell survival to ionizing radiations is a relevant biological endpoint to plan radiotherapy and hadrontherapy treatments since it can be linked to the probability of tumor control. Generally, cell survival is estimated by *in-vitro* measurements of cell-survival curves, which draw the survival probability expressed as a function of the dose delivered by the irradiation facility (see Fig. (1)). To be integrated into a treatment planning system, experimental data have to be accurately reproduced by a model, which can predict survival values at any dose. Within conventional radiotherapy, survival is directly linked to the delivered dose. Consequently, an interpolation by a parametric function of the dose, for which the parameters are fitted to the experimental data, is convenient.

Accordingly, the Linear–Quadratic Model (LQ) proposes a faithful representation with only two free parameters ($\alpha,\beta$):

$$S(D) = e^{-(\alpha.D + \beta.D^2)} \tag{1}$$

This model reproduces in particular the shoulder that can experimentally be observed (see Fig. (1)). In the field of radiotherapy with light ions (hadrontherapy), cell survival depends on dose, but also on ion species and ion energy. A Relative Biological Effectiveness (RBE) ratio was defined to qualify the biological effects of any radiation with regard to another radiation, in general to the *X-rays*. RBE of a given radiation is the ratio of the dose required with *X-rays* over the dose required with the given radiation to get the same biological effect.

The complex behavior of ion-induced biological effects is attributed to the high level of heterogeneity (non-uniformity at microscopic scale) of the dose deposited in an ion track. Indeed, the density of ionized and excited molecules generated along the path can be huge. As a consequence ions may induce complex damages such like double strand breaks or clusters of damage sites in DNA. Such lesions are considered to be very difficult to repair for the cell and to be implied in cell death. In the literature, two approaches have been extensively developed to predict cell inactivation by ion irradiation: the Katz Model and more recently the Local Effect Model (LEM). Both these models are based on a description of track-structure through a radial dose instead of the macroscopic delivered dose, to account for the heterogeneity of energy deposition by swift ions. Radial dose is defined as the averaged local dose deposited by a single ion in an elementary volume expressed as a function of the distance between this volume and the ion trajectory. Radial dose includes ion-charge and ion-velocity parameters, and finally gave to both models the ability to predict important radiobiological features for ion irradiations. It is however an averaged quantity that neglects the stochastic nature of ionizing radiations.

After a brief description of both models, the paper will present a Monte-Carlo simulation of the dose deposited at a microscopic scale and discuss the stochastic effects in terms of dose heterogeneity and track overlapping.

**B THE LEM AND THE KATZ MODELS**

**B.1 THE LOCAL EFFECT MODEL**

To predict cell survival, the Local Effect Model (LEM) [1-2], considers that cell killing arises from the induction of lethal events by the ionizing radiation. Assuming that the distribution of lethal events obeys a Poisson distribution, the probability for the cell to survive reads:

$$S(D) = e^{-N_{lethal}(D)} \quad (2)$$

where $N_{lethal}(D)$ is the mean number of lethal events induced in the cell after a dose *D*. The first key assumption of the LEM is to consider lethal events as point-like events generated by the local dose deposited by the radiation. Thus, the number of lethal events in the cell is the summation of the local lethal events over the cell sensitive volume:

$$N_{lethal}(D) = \iiint_{Sensitive\,Volume} \rho_{lethal}(\boldsymbol{r}) d\boldsymbol{r} \quad (3)$$

where the local density of lethal events is assumed to be a simple function of the local dose *d(r)*:

$$\rho_{lethal}(\mathbf{r}) = \rho_{lethal}(d(\mathbf{r})) \quad (4)$$

In the LEM, the local dose is calculated by cumulative effects, superimposing the local dose deposited by each ion, which is represented by the radial dose $d_R$:

$$d(\boldsymbol{r}) = \sum_i d_R(r_i) \quad (5)$$

where $r_i$ is the radial distance of the point *r* to the trajectory of the $i^{th}$ ion in the transversal plane to the beam axis.

The second key assumption of the LEM consists in extracting the relation between the density of lethal events and the local dose from survival measurements performed with *X-ray* radiation. Indeed, the local dose deposited by *X-ray* radiation is considered as uniform within the cell. Neglecting stochastic effects, it is therefore equal to the macroscopic dose $D$, which is delivered to the sample by the *X-ray* source:

$$d(\mathbf{r}) = D \tag{6}$$

Therefore for *X-ray* irradiation, Eq. (3) becomes simply:

$$N_{lethal}(D) \approx \rho_{lethal}(D).V_{sensitive} \tag{7}$$

According to Eq. (2), $N_{lethal}(D)$, and therefore $\rho_{lethal}(D)$, can be deduced from the measurement of cell survival $S_X(D)$ to *X-ray* irradiation (described by the $\alpha$ and $\beta$ parameters) and from an estimation of the cell sensitive volume $V_{sensitive}$. This latter is assumed to be uniformly distributed over the cell nucleus. The diameter of the sensitive volume depends on the cell and ranges from 5-20 µm. An explicit expression for the average number of lethal events can thus be obtained as:

$$N_{lethal} = \iiint\limits_{Sensitive\ Volume} \frac{-\ln S_X(d(r))}{V_{sensitive}} d\mathbf{r} \tag{8}$$

Practically, *an ion-impact configuration is randomly generated* for a dose $D_{ion}$ with the following procedure 1°) the volume of interest containing the cell is defined large enough such like the energy deposited into the cell by any ion that does not impinge this "interest" volume is insignificant; 2°) the number of ion impacts is generated according a Poisson law for which the parameter, i.e. the mean number of impacts, corresponds to the delivered dose $D_{ion}$; 3°) the impact positions of each ion are generated randomly according to a uniform distribution.

*For this configuration, the local dose is calculated* by Eq. (5): for any point ***r*** of the sensitive volume and for any impinging ion *i*, it is possible to calculate the radial distance $r_i$ and therefore the radial dose $d_R(r_i)$.

*The average number of lethal events* is then deduced from Eq. (8), which gives the probability *S(D$_{ion}$)* for the cell to survive a dose *D$_{ion}$* according to Eq. (2). This process has to be reiterated many times to reduce statistical fluctuations on the predicted survival. Following this protocol, the LEM can in principle predict cell survival to any ion irradiation as soon as the cell sensitive volume and the cell survival to *X-rays* can be experimentally determined. However practically, a set of experimental data performed with ion irradiation is however required to fit a parameter [3] necessary to describe the curve of cell survival to *X-rays* at very high doses (>>10 Grays) since measurements cannot be performed at such doses.

**B.2  THE KATZ MODEL**

Within the Katz Model [4-5], cell inactivation arises from two mechanisms: the *ion-kill* and *γ-kill* modes. Cell-surviving fraction is the product of the surviving fraction associated to both these modes:

$$S = \Pi_i . \Pi_\gamma \tag{9}$$

Both theses modes induce cell killing by inactivating critical biological targets. These targets are characterized by a radius $a_0$ and an inactivation dose $D_0$. Cell killing by *ion-kill* mode is described by the cross-section $\sigma$:

$$\Pi_i = e^{-\sigma F} \tag{10}$$

for which the expression reads:

$$\sigma = \int_0^\infty 2\pi r \left(1 - e^{-\left(\overline{D}(r)/D_o\right)}\right)^m dt, \tag{11}$$

where $F$ is the irradiation fluence and $\overline{D}(r)$ represents the dose deposited in a target for which the center stands at a distance $r$ ($t$ within the author notation) from the ion path. The exponent $m$ means that $m$ targets have to be inactivated to induce cell death. For a fixed ion species, $\sigma$ increases with ion LET to a plateau $\sigma_0$ and then may decrease.

The survival fraction associated to *γ-kill* mode is given by:

$$\Pi_\gamma = 1 - \left(1 - e^{-\left(D_\gamma/D_o\right)}\right)^m \tag{12}$$

$D_\gamma$ is the contribution to the delivered dose $D$ which corresponds to $\gamma$-kill mode. It is given by:

$$D_\gamma = (1-P)D \tag{13}$$

where the fraction $P$ is given by:

$$P = \frac{\sigma}{\sigma_0} \tag{14}$$

In the Katz Model, the description of the track structure appears through the calculation of the dose $\overline{D}(r)$ deposited into the sensitive targets. As for the LEM model, the track-structure is described by the radial dose.

As already mentioned in the introduction, radial dose is an averaged quantity, which therefore does not take fully into account the stochastic nature of ionizing radiations. To evaluate the impact of such an approximation, it seems relevant to simulate with a Monte Carlo simulation the local dose for the LEM model and the dose deposited into targets for the Katz Model.

### C MATERIAL AND METHODS

The calculation of dose (or energy) depositions into targets by Monte-Carlo simulation has ever been undertaken by various authors for other purposes [6-7] and with various methods. Here, we propose a method that matches with the integration of Eq. (8) into the LEM. Precisely, a water sample was divided into cubes (a cube mesh) whose size corresponds to the spatial extension of the lethal events for the LEM or to the target diameter ($2a_0$) for the Katz Model. While the target radius $a_0$ is clearly defined in the Katz Model to 500-1000 nm, the literature that describes the LEM do not fix the scale of locality. Instead lethal events are considered as point-like events and by using a radial dose the question can be averted. However to calculate a microscopic dose, i.e. the specific energy within microdosimetry terminology, the target extension has to be set. Lethal-event extension is necessarily larger than the atomic scale. It may be of the order of double-strand-break extension, which is typical less than 20 pairs of bases (~6 nm) [8-9]. One has also to consider that damage may be created by indirect actions. Therefore a damage site can be induced by a water radical produced elsewhere, and the diffusion distance is generally assumed to be of the order of few nanometers [8]. Finally, the mesh cube size was set to 10 nm to evaluate the stochastic effects in the LEM.

Our Monte-Carlo simulation consists in following the incident particles and all the produced electrons in the induced electronic cascades. The electrons are followed until their energy becomes lower than a cut-off energy set to 33 meV (300K). The result of the simulation is a spatial distribution of low-energy electrons, and ionized or excited water molecules. The dose deposited into a cube is calculated by summing the energy of all species in this cube at the end of the simulation. We would like to emphasize that the total stored energy represents about 85% of the energy transferred to the water sample directly by the radiation since a part of the energy is converted into target heating: below the threshold energy for water-molecule excitation and ionization, electrons lose their energy by phonon creation. Moreover, ionized molecules can be neutralized by electron capture and excited molecules can go back to ground state by non radiative processes.

For ions, all details of our simulation can be found in [10-11]. For *X-ray* irradiation, we simulated the irradiation by 1.3 MeV photons, which correspond to a major component of the spectrum that characterizes the *γ-rays* generated by a $^{60}$Co source. At this energy, Compton interaction dominates and most of the photon interactions eject Compton electrons. The Compton-electron transport is simulated with the same code used [10-11] for ion irradiation. In both irradiation modalities, we applied to the irradiated sample periodical boundary conditions to avoid edge effects and to ensure equilibrium of charged particles. The sample was chosen either to mimic cell nucleus ($10 \times 10 \times 10 \mu m^3$) or large enough to reduce statistical errors in histograms ($50 \times 50 \times 10 \mu m^3$).

**D RESULTS**

*Heterogeneities*

We simulated the irradiation of a 50x50x10 µm$^3$ sample at a dose of 1 Gray with $^{60}$Co *γ-rays*, H[10MeV] and C[10MeV/n] ions. The mesh cube size was set to 10 nm. The first observation we made is that most of the cubes do not receive any energy transfers, whatever the irradiation. More precisely, the probability for a cube to receive energy is 5.2 10$^{-4}$ for $^{60}$Co *γ-rays*, 3.7 10$^{-4}$ for H[10MeV] and 2.0 10$^{-4}$ for C[10MeV/n]. We can therefore conclude that, even for *X-ray* irradiation, the local dose is non-uniformly spread over the sample. Fig. (2) compares histograms, which represent the probability for a cube to receive a given local dose, calculated for the three different projectiles. We observed that the distribution of local dose is broad. It varies from 10 Gray to a few 10$^5$ Grays, whatever the projectile. Below 10$^4$ Grays, the histograms are similar and reveals the strong heterogeneity, which is a common feature of all 3 radiations. To clarify this observation we will focus on the differences that can be observed at very high local doses between low- and high-LET radiations. One can expect the ion track core to play a significant role since the density of ionization can be huge in this region for such high-LET particles. We defined an "ultra-track" ion histogram by disregarding the events in the track core. Fig. (3) compares the histograms of local doses calculated for an irradiation dose of 1 Gray with $^{60}$Co *γ-rays* and C[77MeV/n]. On this figure was also plotted the histogram of C[77MeV/n] after suppressing the track–core contribution ($R_{core}$=10nm). As expected, the structure standing above $10^5$Gy is removed as soon as the track core is suppressed. This structure at high local doses is therefore specific to high-LET ions and may significantly contribute to the high values of RBE, at least in the framework of models based on local effects. To compare more accurately the ultra-track histogram with the histogram of local dose simulated for $^{60}$Co *γ-ray* radiation, we normalized it to a delivered dose of 1 Gray and plotted the normalized histogram on Fig. (3). The global agreement suggests that ion-induced fast δ-electrons generate a distribution of local doses very similar to the *γ-rays* of $^{60}$Co. The common features of all these histograms are likely related to the pattern of these fast δ-electron collisions with water molecules. Fig. 4 compares the histogram of the energy deposited into cubes to the energy distribution of isolated events (ionization, excitation, electron attachment…). The latter histogram is equivalent to the histogram of energy deposition calculated with a tiny cube size. It has been arbitrarily normalized for the sake of easy comparison. For these calculations the delivered dose was 1 Gray and the projectile was a C[10MeV/n] ion beam. One can observe that a part of the features in the local-dose histogram can be attributed to the generation of elementary events. The region below 33 meV, which is the cut-off energy of the electron-cascades, has to be associated to the thermalized electrons. Between 33 meV and 6 eV, stand the geminate-recombination events. Such recombination events generally evolve into excited water molecules. The stored energy for this class of events is the kinetic energy of the ejected electrons before recombination. The region above 6 eV is associated to electron attachments (6-12.4 eV), to water-molecule excited states (8.2 eV and 10 eV) and to the serial of ionized and multi-ionized states. In particular the multi-ionization with single and double ionization in K-shell can be clearly distinguished. Although the region of thermalized electrons and the energy structure for attachment, excitation and ionization can be clearly identified in the local-dose histograms, many histogram features arise from a combination of numerous events. Fig. (5) decomposes the histograms of local dose according to the number $N$ of events that occur in cubes for an irradiation of 1 Gray with C[10MeV/n]. Clusters of events (large N) strongly mark up the histograms. In particular, in the region of doses larger than $\sim 10^4$ grays the number of events is larger than 10 per cube. For *X-ray* irradiations, high-dose deposition is likely due to ionization clusters generated by low-energy electrons. For ion irradiations, track core is the major contribution to these high-local-dose depositions.

From this analysis, we can conclude that local dose cannot be considered as uniform neither for ion irradiation nor for *X-ray* irradiation. Heterogeneities arise from the energy spectrum of elementary events and from the combination of events. Eq. (8), which links the averaged number of lethal events to the local dose in the LEM, is therefore questionable. Indeed, this latter expression is based on the assumption that fluctuations of local doses within a cell nucleus can be neglected for *X-ray* irradiation, while we have shown that they are actually huge. But, it is important to emphasize that these heterogeneities, which are neglected for *X-rays*, are considered to be at the origin of the radiobiological efficiency of ions.

*Track overlapping*

Track overlapping plays a significant role in the predictions of both the Katz Model and the LEM. In the LEM, track overlapping induces cell-killing with a higher efficiency than independent tracks would do. Indeed, due to the shoulder in the curves of cell survival to *X-rays*, the local density of lethal events increases non-linearly with the local dose. As a consequence, the lethal effect induced by the superimposition of the local dose generated by two independent ion tracks is larger than the addition of their individual effects

$$\rho_{lethal}(d_1 + d_2) \geq \rho_{lethal}(d_1) + \rho_{lethal}(d_2) \tag{15}$$

The left-hand-side part of this equation appears as high order terms in delivered dose in Eq. (3) and, through Eq. (2), is responsible for the apparition of shoulders in the LEM.

Generally speaking, the lower the LET is, the more significant the overlapping is. Indeed, for a fixed dose, the number of impacts decreases as LET increases. Moreover, for an irradiation with ions of a given LET, decreasing ion velocity decreases the spatial extension of the tracks and therefore decreases the probability for overlapping.

From these considerations, one could understand why a shoulder can be observed for low-LET ions and why it disappears for ions at Bragg peak (low velocity and high LET).

To evaluate how stochastic effects might modify this overlapping scheme, we simulated the local-dose deposition for *X-ray* irradiations with 1 Gray and 5 Grays (see Fig. (6)). At such doses and for *X-ray* irradiation, an effect of overlapping is expected to be significant since shoulders appear within this range of doses. Instead, we observe that both histograms are identical except for the factor of 5 in histogram amplitude, which corresponds to the dose ratio. In other words, at local scale, two incident particles cannot significantly contribute to the same local site. In fact, increasing the dose simply increases the number of hit cubes.

This observation can be mathematically written as

$$h_D(d_k) = h_1(d_k).D \quad \forall d_k \neq 0 \qquad (16)$$

where $h_D$ (resp. $h_1$) is the histogram of local dose $d_k$ for an irradiation dose of $D$ (resp. 1 Gray). Such a relation can be conveniently inserted into Eq. (3) of the LEM after rewriting the latter:

$$N_{lethal} = \int_{d'} dd' \rho_{lethal}(d') \iiint_{SensitiveVolume} d\mathbf{r} \delta(d'-d(\mathbf{r})) \qquad (17)$$

The integration over the sensitive volume represents the volume associated to a local dose $d'$:

$$\frac{d\mathbf{r}}{dd'} = \iiint_{SensitiveVolume} d\mathbf{r} \delta(d'-d(\mathbf{r})) \qquad (18)$$

It is simply related to the histogram of local doses by normalization to the sensitive volume $V_{Sensitive}$:

$$h(d') = \frac{d\mathbf{r}}{dd'} \Big/ V_{Sensitive} \qquad (19)$$

One gets then:

$$N_{lethal} = V_{Sensitive} \int_{d'} dd' \rho_{lethal}(d') h(d') \qquad (20)$$

It is therefore possible to introduce Eq. (16) for a delivered dose $D$:

$$N_{lethal}(D) = V_{Sensitive} \cdot \int_{d'} dd' \rho_{lethal}(d') D.h_1(d') \qquad (21)$$

Defining then a constant $\alpha$ by:

$$\alpha = N_{lethal}(D = 1\,Gy) \qquad (22)$$

we can deduce from Eq. (2) that cell surviving fraction at a dose $D$ obeys to:

$$S(D) = e^{-\alpha D} \tag{23}$$

This clearly shows that a pure local effect theory cannot predict shoulder in cell survival. This conclusion is robust: we have verified that Eq. (16), which is necessarily valid for cube length smaller than 10 nm, is still valid for cube length as large as 100 nm. All these conclusions raise questions on the physical meaning of the shoulders predicted by the LEM. The apparition of a shoulder in the LEM comes from the introduction of an averaged quantity, namely the radial dose, to represent the track structure. Indeed, as it is illustrated by Fig. (7), at the scale of local events, overlapping of tracks is actually very scarce for doses lower than 10 Grays. Superimposing radial dose comes down to superimpose events that occurred apart. Finally by superimposing averaged quantities, the LEM introduces artificial non-local effects and therefore artificial shoulders.

Within the Katz Model overlapping acts through  mode, while *ion-kill* mode is restricted to intra-track effects. Within both modes, cell-killing is induced by the same process, the inactivation of m targets characterized by the same geometry and by the same inactivation dose. However the delivered dose is shared into a *γ-kill* mode contribution and an *ion-kill* mode contribution. Despite the repartition of dose to deal with overlapping, the formulation of the Katz Model raises up many questions. Fig. (8) compares histograms of dose calculated for an irradiation of C[77MeV/n] with 1 Gray and 5 Grays. For these calculations, the mesh size was set to 1 µm to represent the geometry of the biological targets in the Katz Model. One can clearly observe that overlapping deeply modifies the histogram structures. For such a spatial extension of biological targets, overlapping cannot be neglected but instead dominates. Such an observation supports the idea of introducing an inter-track mode, but questions the expression of cell survival for *ion-kill* mode. A description of cell survival by a cross-section may be valid for very low fluences. Indeed, while overlapping remains low, the number of biological targets that are inactivated by intra-track process increases proportionally to the fluence. Surviving fraction therefore decreases exponentially with the dose. However for dose values of a few Grays, overlapping tends to be dominant and the probability for a target to receive energy from one impact only (intra-track process) decreases with the dose. Cell-killing by *ion-kill* mode is therefore overestimated and Eq. (10) underestimates the fraction of cell survival to *ion-kill* mode.

The calculation of cell survival to $\gamma$-*kill* mode has also to be discussed through the estimation of dose which is attributed to $\gamma$-*kill* mode. According to Eq. (13), this $\gamma$-*kill* dose is fully determined by the fraction P, which represents the fraction of dose associated to *ion-kill* mode. From our analysis, it is clear that the fraction of dose associated to *ion-kill* mode should decrease with the dose: at very low doses, overlapping is insignificant and inactivation of targets issues from intra-track process. P is expected to be of the order of 100%. Instead, at a dose value of a few Grays, overlapping, and therefore inter-track process, dominates. However, according to Eq. (14), the fraction P is set independent of the delivered dose.

**E CONCLUSIONS**

The Katz Model, and later on the Local Effect Model, proposed a scenario, based on track-structure, that explains the main features of cell survival to ion irradiation. By means of a restricted number of free parameters, both theses models predict, for cell surviving fraction, values relevant enough to be integrated into hadrontherapy treatment plans. Probably for the sake of practical reasons, theses track-structure models disregarded stochastic effects. Both models represent the track-structure by a radial dose, which is an averaged quantity. However we have shown that stochastic effects cannot be neglected and that working with averaged quantities is questionable. Generally speaking, for huge fluctuations and for a non-linear function, the function average differs from the function of the average. In the LEM the non-uniformities of local-dose are considered for ion irradiations but strangely washed-out for low-LET irradiations. Moreover we have shown that the shoulders predicted by the LEM indeed arise from artificial non-local effects. In Katz Model, although the definition of a cross-section for ion-induced cell killing is relevant, its application for dose values larger than 1 Gray is questionable because of the spatial extension of the biological targets. Moreover the fraction of dose attributed to *ion-kill* mode and *γ-kill* mode is strangely assumed to be dose independent. One instead may expect overlapping, and therefore inter-track processes, to increase with dose while intra-track processes should decrease. The story of track-structure model has not reached its end yet. A more fundamental knowledge including the stochastic nature of ionizing radiation is required. Finally, new simple model for RBE prediction should be built in such a way to be consistent with the observed statistical properties of dose deposition

ACKNOWLEDGEMENTS

This work has been supported by the INCa institute (Institut National Du Cancer) and CPER Rhône-Alpes.

REFERENCES

[1]     M. Scholz and G. Kraft. Track structure and the calculation of biological effects of heavy charged particles. *Adv. Space. Res.*, 18(1/2):5–14, 1996.

[2]     M. Scholz, A.M. Kellerer, W. Kraft-Weyrather, and G. Kraft. Computation of cell survival in heavy ion beams for therapy the model and its approximation. *Radiat. Environ. Biophys.*, 36:59–66, 1997.

[3]     M. Beuve, G. Alphonse, M. Maalouf, A. Colliaux, P. Battiston-Montagne, P. Jalade, E. Balanzat, A. Demeyer, M. Bajard, and C. Rodriguez-Lafrasse. Radiobiologic parameters and local effect model predictions for head-and-neck squamous cell carcinomas exposed to high linear energy transfer ions. *Int. J. Rad. Onc. Bio. Phys.*, 71(2):635–42, Jun 2008.

[4]     R. Katz, B. Ackerson, M. Homayoonfar, and S. C. Sharma. Inactivation of cells by heavy ion bombardment. *Radiat. Res.*, 47(2):402–425, Aug. 1971.

[5]     Katz R., Zachariah R, Cucinotta FA, and Zhang C. Survey of cellular radiosensitivity parameters. *Radiat. Res.*, 140(3):356–65, Dec. 1994.

[6]     Nikjoo H, O'Neill P, Goodhead DT, Terrissol M. Computational modelling of low-energy electron-induced DNA damage by early physical and chemical events. *Int. J. Radiat. Biol.,* 71(5):467-83, May 1997.

[7]     Champion C.; L'Hoir A.; Politis M.F.; Chetioui A.; Fayard B.; Touati A. Monte-Carlo simulation of ion track structure in water: ionization clusters and biological effectiveness, *Nucl. Instr. and Meth. in Phys. Res. Section B*, 146(1), 533-540(8), Dec. 1998.

[8]     T. Elsasser and M. Scholz. Cluster effects within the local effect model. *Radiat. Res.*, 167(3):319–29, March 2007.

[9]     M. Scholz and G. Kraft. The physical and radiobiological basis of the local effect model: a response to the commentary by r. katz. *Radiat. Res.*, 161:612 – 620, 2004.

[10]    B. Gervais, M. Beuve, G. H. Olivera, and M. E. Galassi. Numerical simulation of multiple ionization and high LET effects in liquid water radiolysis. *Radiat. Phys. Chem.*, 75:493–513, April 2006.

[11]    B. Gervais, M. Beuve, G. H. Olivera, M. E. Galassi, and R. D. Rivarola. Production of HO2 and O2 by multiple ionization in water radiolysis by swift carbon ions. *Chem. Phys. Lett.*, 410:330–334, July 2005.

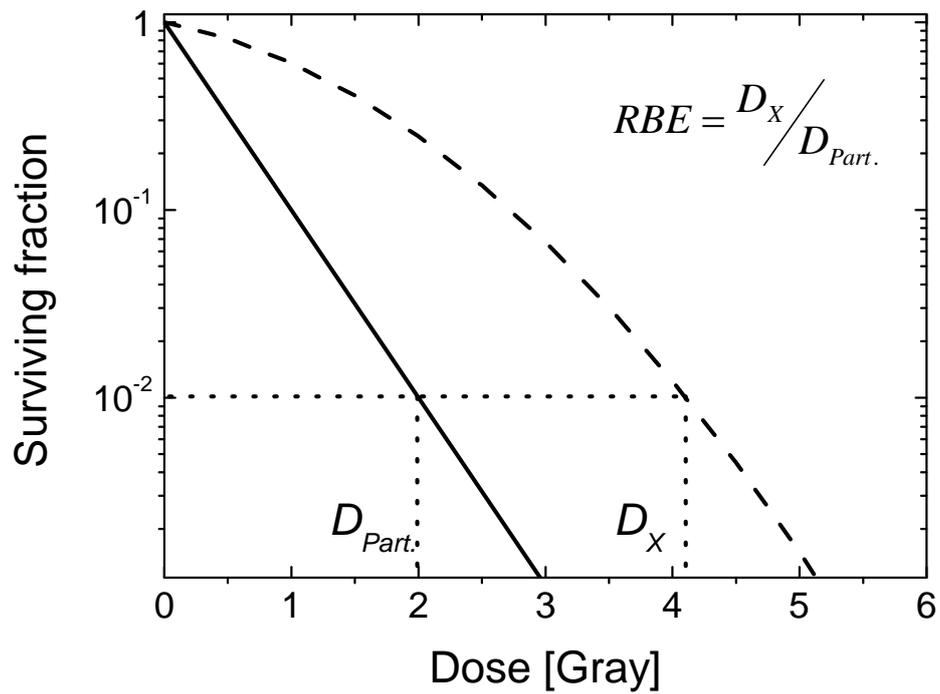

Figure 1: Definition of RBE: dashed line (resp. solid line) represents the fraction of cells that survive an *X-ray* (resp. light ion) irradiation versus the delivered dose. A shoulder shape is visible for the *X-ray* curve.

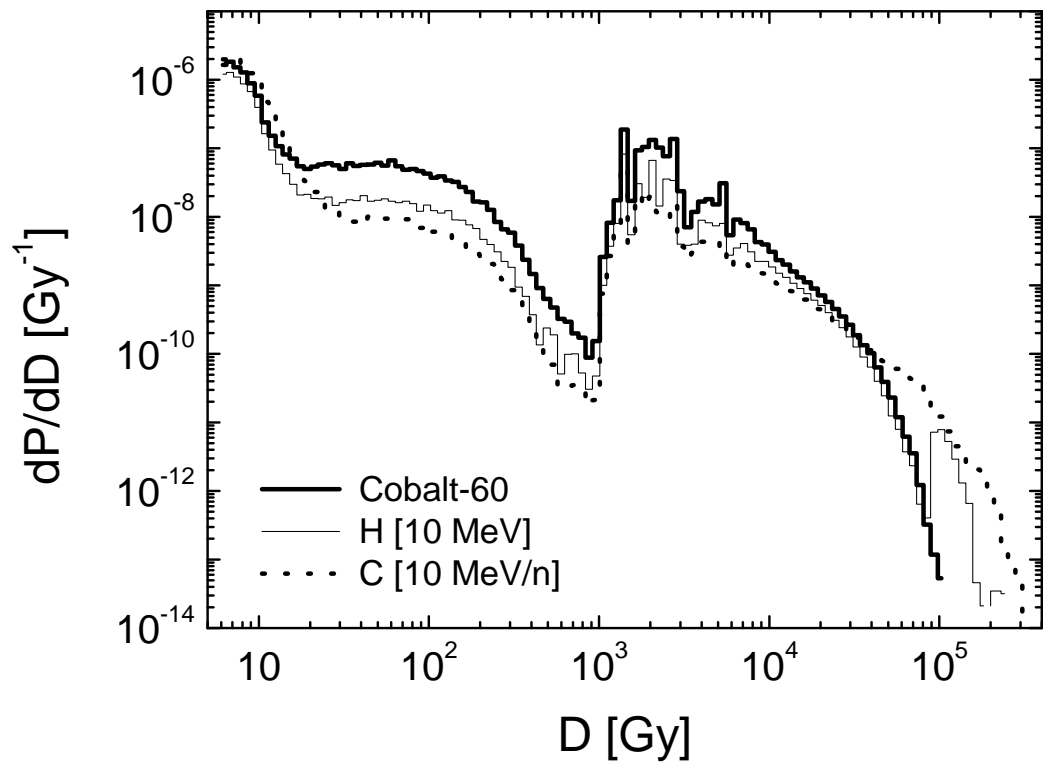

Figure 2: Histogram of local dose calculated for a water sample of 50x50x10 µm³ irradiated at a dose of 1 Gray with a beam of $^{60}$Co *γ-rays*, H[10MeV] and C[10MeV/n]. The mesh resolution is 10 nm.

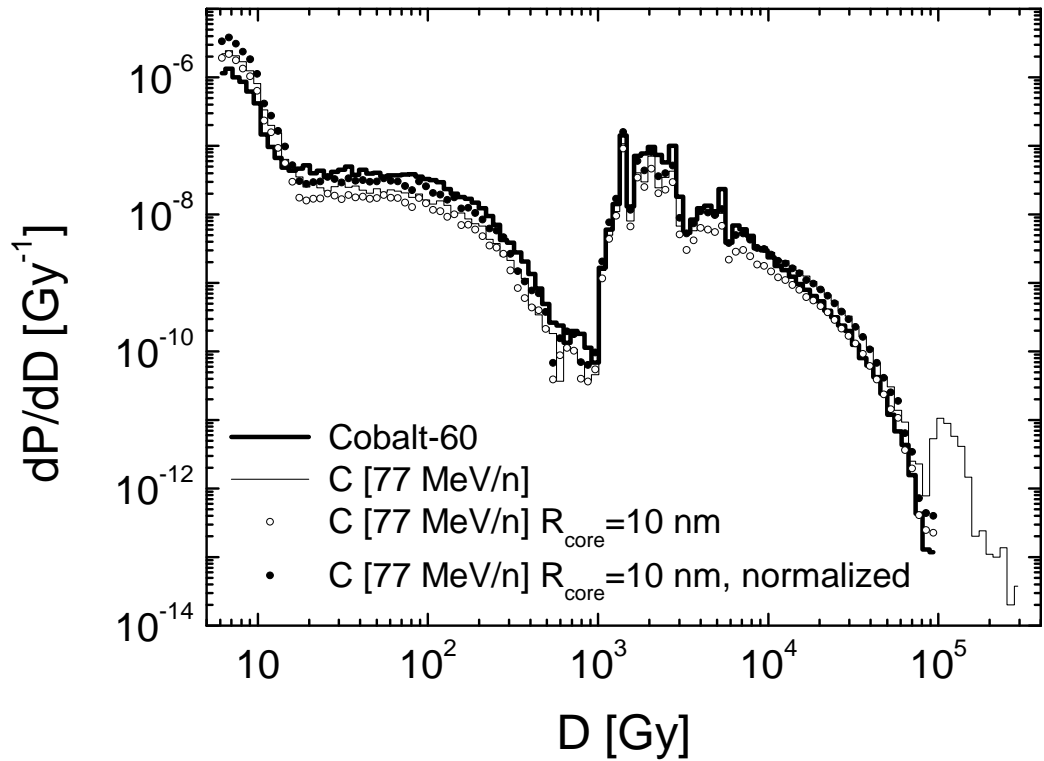

Figure 3: Histogram of local dose calculated for a water sample of 10x10x10 µm$^3$ irradiated at a dose of 1 Gray with a beam of $^{60}$Co $\gamma$-rays and a beam of C[77MeV/n]. The mesh resolution is 10 nm. The curves labeled with "$R_{core}$" represent the contribution of the ultra-track (see text). The last curve was normalized to get an average dose of 1 Gray despite of the suppression of the track-core contribution.

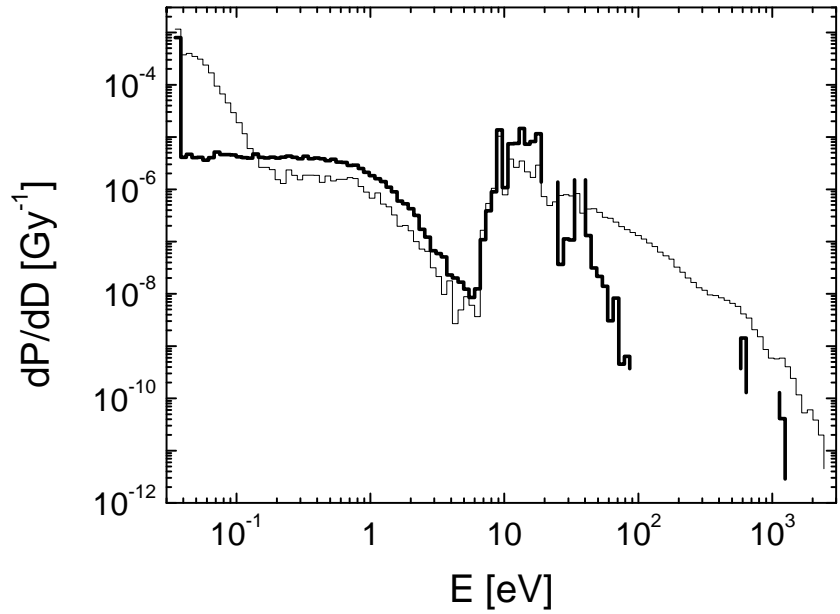

Fig 4: Histogram of energy transfers for C[10MeV/n] irradiation. Thin line: energy transferred to the 10x10x10 nm$^3$ cubes of a 10x10x10 µm$^3$ sample irradiated at a dose of 1 Gray. Thick line: energy of the individual events generated by the irradiation. This latter curve was arbitrarily normalized for the sake of easy comparison.

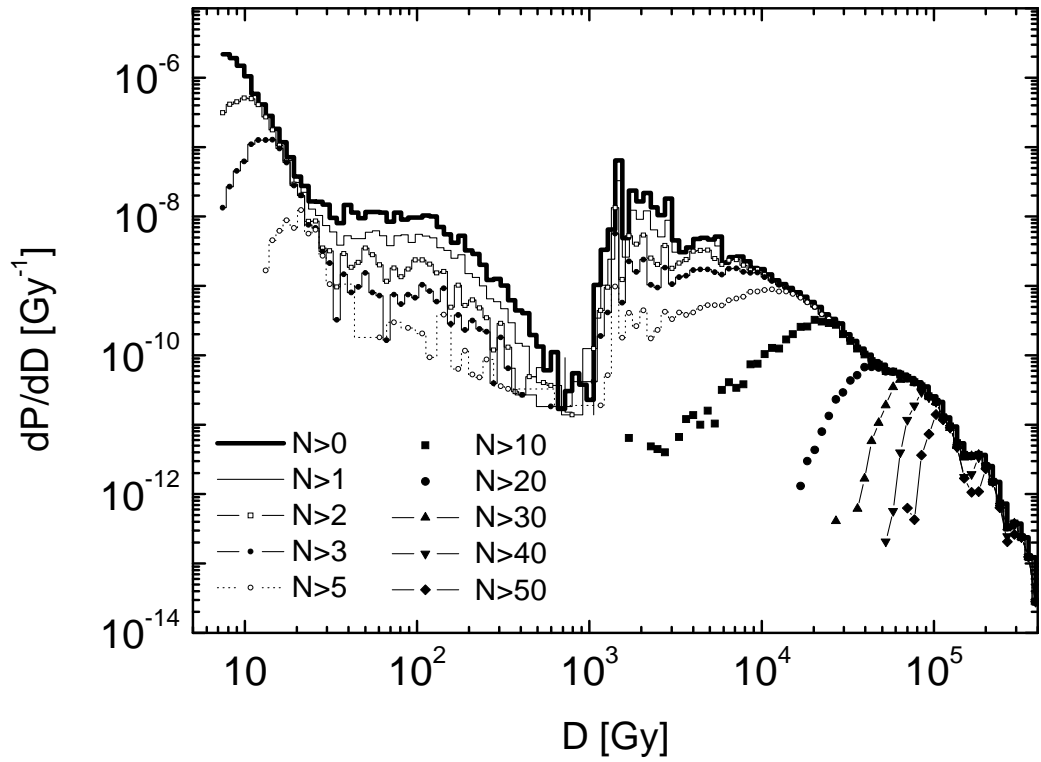

Fig. 5 Histogram of local dose calculated for a sample of $10 \times 10 \times 10\ \mu m^3$ irradiated at a dose of 1 Gray with C[10MeV/n]. The mesh resolution is 10 nm. The curves labeled with "N>$m$" refer to histograms calculated by only considering the cubes that receive a number N of events per cube larger than $m$.

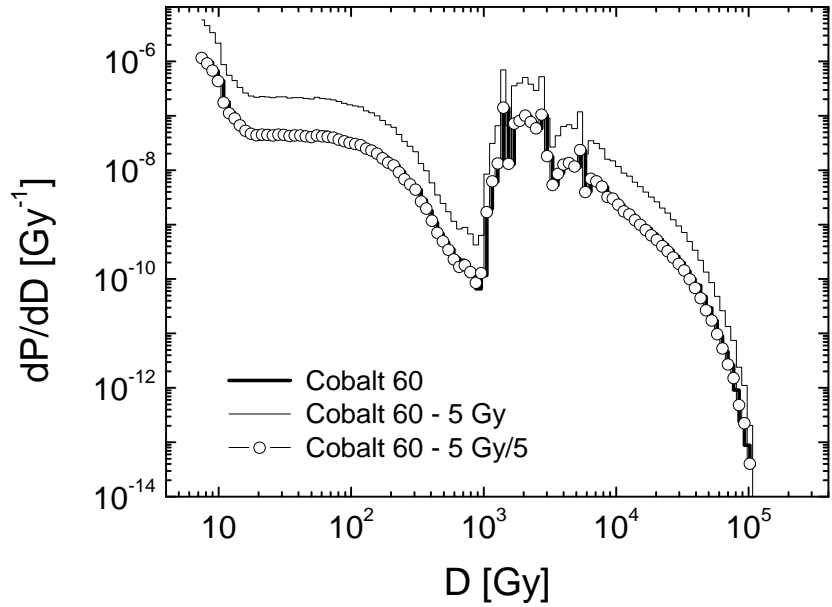

Figure 6: Histogram of local dose calculated for a sample of 10x10x10 μm$^3$ irradiated with a dose of 1 Gray and 5 Grays of $^{60}$Co *γ-rays*. The mesh resolution is 10 nm. The curve labeled with "Cobalt 60-5Gy/5" represents the histogram calculated for 5 Grays and divided by a factor 5.

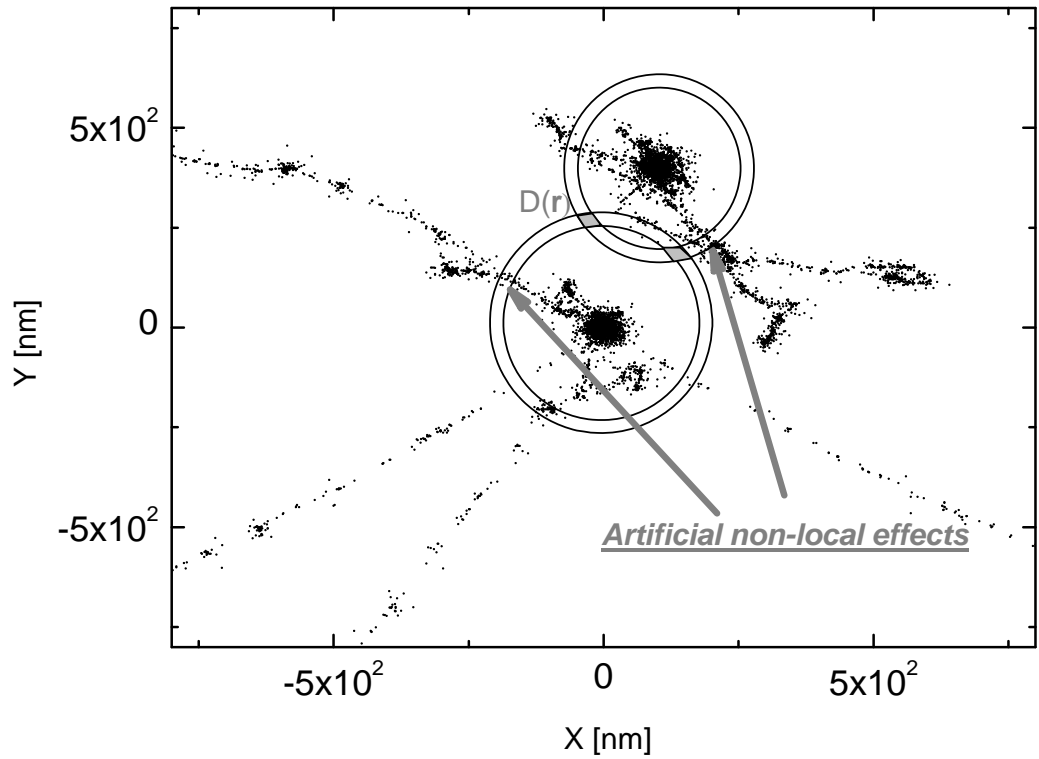

Figure 7: Overlapping of two ion tracks. The two couples of circles illustrate the calculation of local dose by superimposition of the radial dose. The two sets of points correspond to the projection of two track segments simulated by Monte-Carlo simulation. The two arrows point to two clusters of events that are considered to superimpose in the framework of the LEM although they are clearly apart.

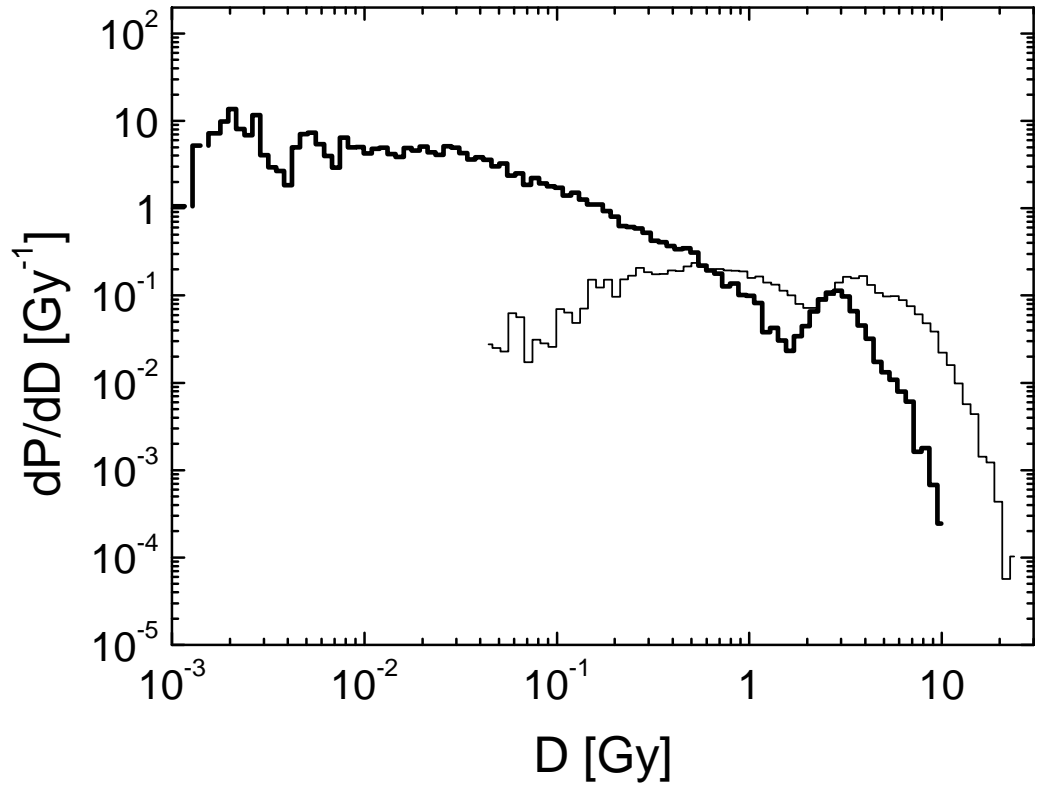

Figure 8: Histograms of dose calculated for a water sample of 30x30x10 μm$^3$ irradiated at a dose of 1 Gray (thick line) and 5 Grays (thin line) with a beam of C[77 MeV/n]. The mesh resolution is 1000 nm.